\documentclass{eas}
\usepackage{graphicx}
\newcommand{\teff}{\mbox{${T}_{\rm eff}$}}

\newcommand{\msun}{\mbox{${\rm M}_{\odot}$}}

\begin{document}

\title{Comparisons for ESTA-Task3: CLES and CESAM} 
\author{J. Montalb\'an}\address{Institut d'Astrophysique et Geophysique, Universit\'e de Li\`ege}
\author{S. Th\'eado$^1$}
\author{Y. Lebreton}\address{Observatoire de Paris-Meudon, GEPI}
\begin{abstract}
We present the results of comparing three different implementations of 
the microscopic diffusion process in the stellar evolution codes CESAM and
CLES. For each  of these implementations we computed models of 1.0,
1.2 and 1.3~\msun. We analyse the differences in their internal structure
at three selected  evolutionary stages, as well as the variations of  
helium abundance and depth of the stellar convective envelope. The origin of
these differences and their effects on the seismic properties of the
models are also considered.
\end{abstract}
\maketitle
\section{Introduction}

As reported in Monteiro {\em et al.} (2006, and references therein), 
 the stellar models
provided by the codes CESAM2k (Morel, 1997) and CLES (Scuflaire {\em et al.}, 2007a)
for a given set of standard input physics, differ by less than 0.5\%. 
At variance with previous comparisons, in this new ESTA-TASK3 we deal with
stellar models that include microscopic diffusion.
The treatment of the  microscopic diffusion process in the evolution codes
we test here, is not exactly the same.
 CLES code computes the diffusion coefficients by solving the 
Burgers' equations (Burgers, 1969) with the formalism developed in 
Thoul {\em et al.} (1994, thereafter TBL94). CESAM2k  provides  two 
approaches to compute diffusion velocities: one  (which we will call  CESAM2k~MP) 
is based on Michaud \& Proffitt (1993) approximation, 
the other  (hereafter CESAM2k~B)  is based on the Bugers' formalism, with  
collision integrals derived from Paquette {\em et al.} (1986).

We will compare three sets of models Task3.1 (1.0~\msun), Task3.2 (1.2~\msun)
and Task3.3 (1.3~\msun), whose  input parameters and physics specifications
are described in Lebreton (2007).  
In the next sections we will present the results of comparing the stellar  
models that were calculated by CLES, CESAM2k~MP and  CESAM2k~B  
for the three sets of models, and we  try to find out  the reason of 
the differences we get.
\begin{figure}[t]
\begin{center}
\includegraphics[scale=0.45]{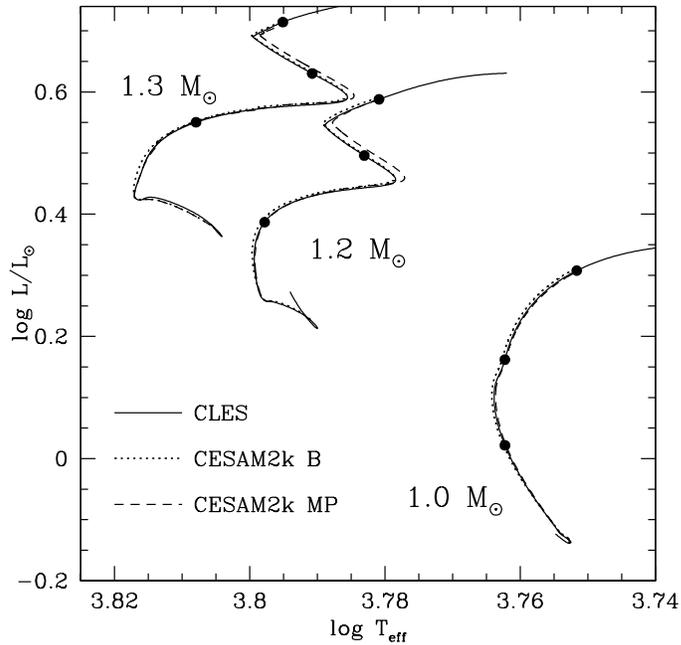}
\caption{Evolutionary tracks corresponding to 1.0, 1.2 and 1.3~\msun. Solid lines: CLES models 
with TBL94 diffusion algorithm.
 Dotted-lines: CESAM2k with Burgers' equations and Paquette {\em et al.} (1986) collision
integrals. Dashed-lines: CESAM2k with Michaud \& Proffitt (1993) approach. The full-dots
along the evolutionary tracks correspond to models with $X_{\rm c}$=0.35 and 0.01, and with
a He-core mass $M_{\rm c}^{\rm He}=0.05\,M_{\star}$.}
\label{fighr}
\end{center}
\end{figure}

\section{Stellar structure and evolution}

For each Task3 we select three evolutionary stages: A: a main sequence stage with a central 
hydrogen content $X_{\rm c}=0.35$; B: a stage close to the core hydrogen exhaustion 
$X_{\rm c}=0.01$, and C: a post-main sequence stage in which the mass of the helium
core (defined as the central region where the hydrogen mass fraction is $X\leq0.01$)
is $M_{\rm c}^{\rm He}=0.05\,M_{\ast}$. 
CESAM stellar models have a number of mesh points between 2700 and 3100, depending on the 
evolutionary stage, while CLES models have about 2400 mesh points. Moreover, for
 all the models considered in these comparisons, the stellar structure ends at $T=$\teff. 
Concerning the time step, both codes make from 1000 to 1500 (depending on the
stellar mass) time steps to reach  stage C, and 
the specifications for the stages  A, B and C
are achieved with a precision better than $1\times 10^{-4}$.

\begin{figure}[t]
\begin{center}
\includegraphics[width=\textwidth]{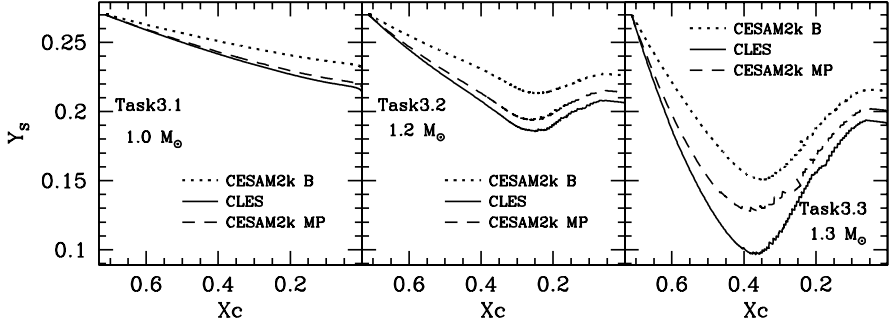}
\vspace*{-1cm}
\caption{Evolution of helium content in the convective envelope ($Y_{\rm S}$ {\em vs.} $X_{\rm c}$)
for 1.0~\msun\ (left panel), 1.2~\msun\ (central panel) and 1.3~\msun\ (right panel).}
\vspace*{-0.7cm} 
\label{figys}
\end{center}
\end{figure}

 Fig.~\ref{fighr} displays, for each  microscopic diffusion implementation,
the  evolutionary tracks for   Task3.1, 3.2 and 3.3,  and the HR diagram location of the 
target models A, B and C. For each stellar mass, the main sequence computed with CESAM2k~B is slightly
hotter ($\sim 0.1$\% for task3.1, to 0.3\% for task3.2 and 3.3)  than those calculated by
CESAM2k~MP and  CLES. 
Furthermore, CLES and CESAM2k~MP models are quite close ($\Delta R/R < 0.3$\%, and $\Delta L/L <0.4$\%)
with the exception of models in the second overall contraction phase, for which the differences
can reach 1\% in the stellar radius and 0.5\% in luminosity.
The fact that  CESAM2k~B models are hotter than CESAM2k~MP and CLES ones 
 could  suggest that the outer layer opacity for the former is lower than for 
the latter because of a different content of hydrogen in their  convective envelope.
The evolution of the helium abundance in the stellar convective envelope ($Y_{\rm S}$) is an eloquent
indicator of the microscopic diffusion effects. Fig.~\ref{figys} shows, 
for each considered stellar mass and
diffusion treatment, the variation of $Y_{\rm S}$ as the central hydrogen content $X_{\rm c}$ decreases,
and reveals that the diffusion efficiency in CLES  is always larger than in CESAM:
about 8, 10 and 20\% larger than  in CESAM2k~MP for 1.0, 1.2 and 1.3~\msun\ respectively,
and 40\%  larger than in CESAM2k~B for all  stellar masses under consideration.

The irregular behaviour of $Y_{\rm S}$ {\em vs.} $X_{\rm c}$ curves for task3.2 and 3.3, is 
a consequence of a semiconvection phenomenon that appears below the convective envelope
and,  the longer main sequence for CESAM2k~MP models is probably due to semiconvection at the
border of the convective core (see next section). 

\begin{figure}
\begin{center}
\vspace*{-0.5cm}
\includegraphics[angle=-90,width=\textwidth]{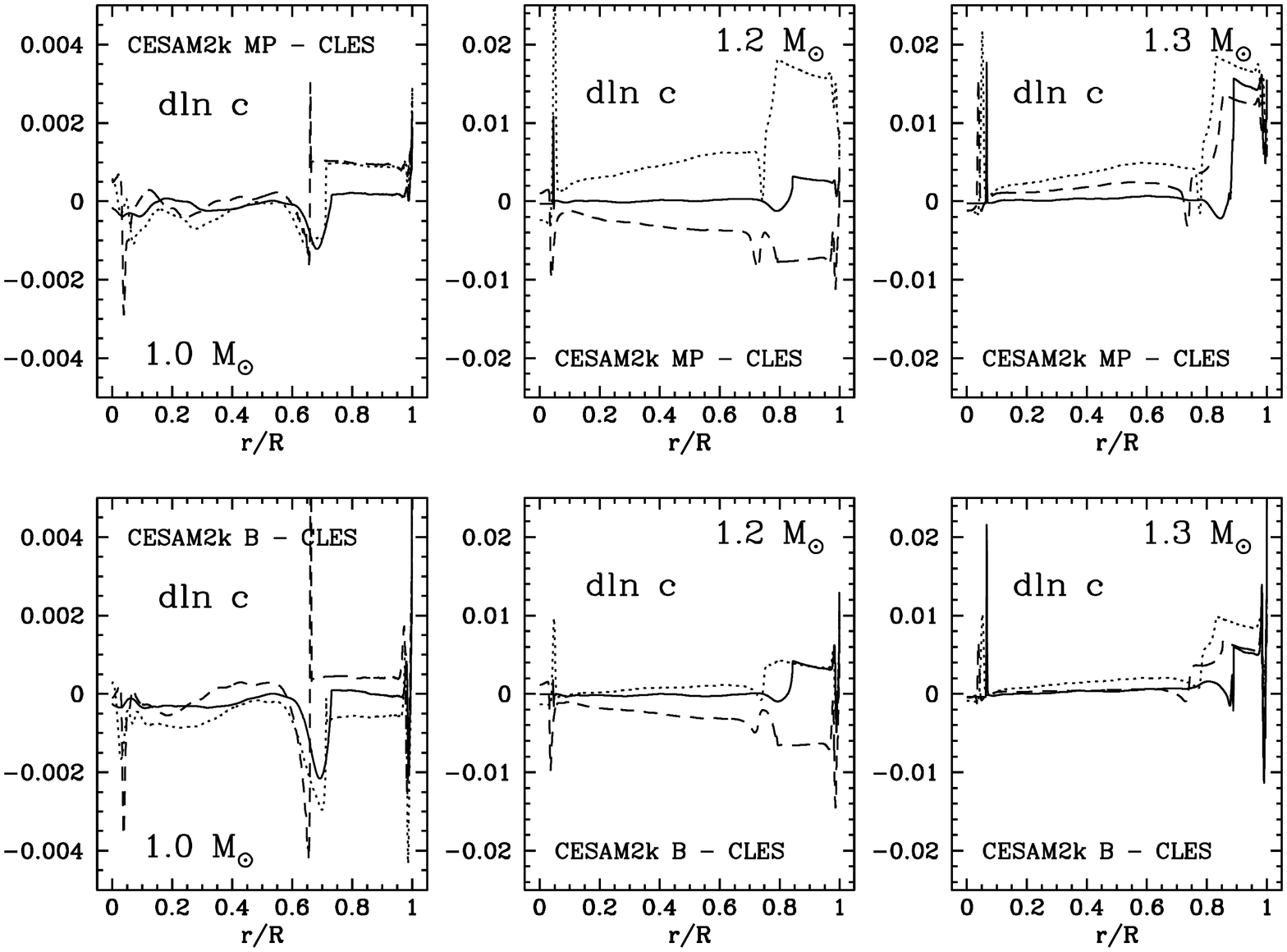}
\includegraphics[angle=-90,width=\textwidth]{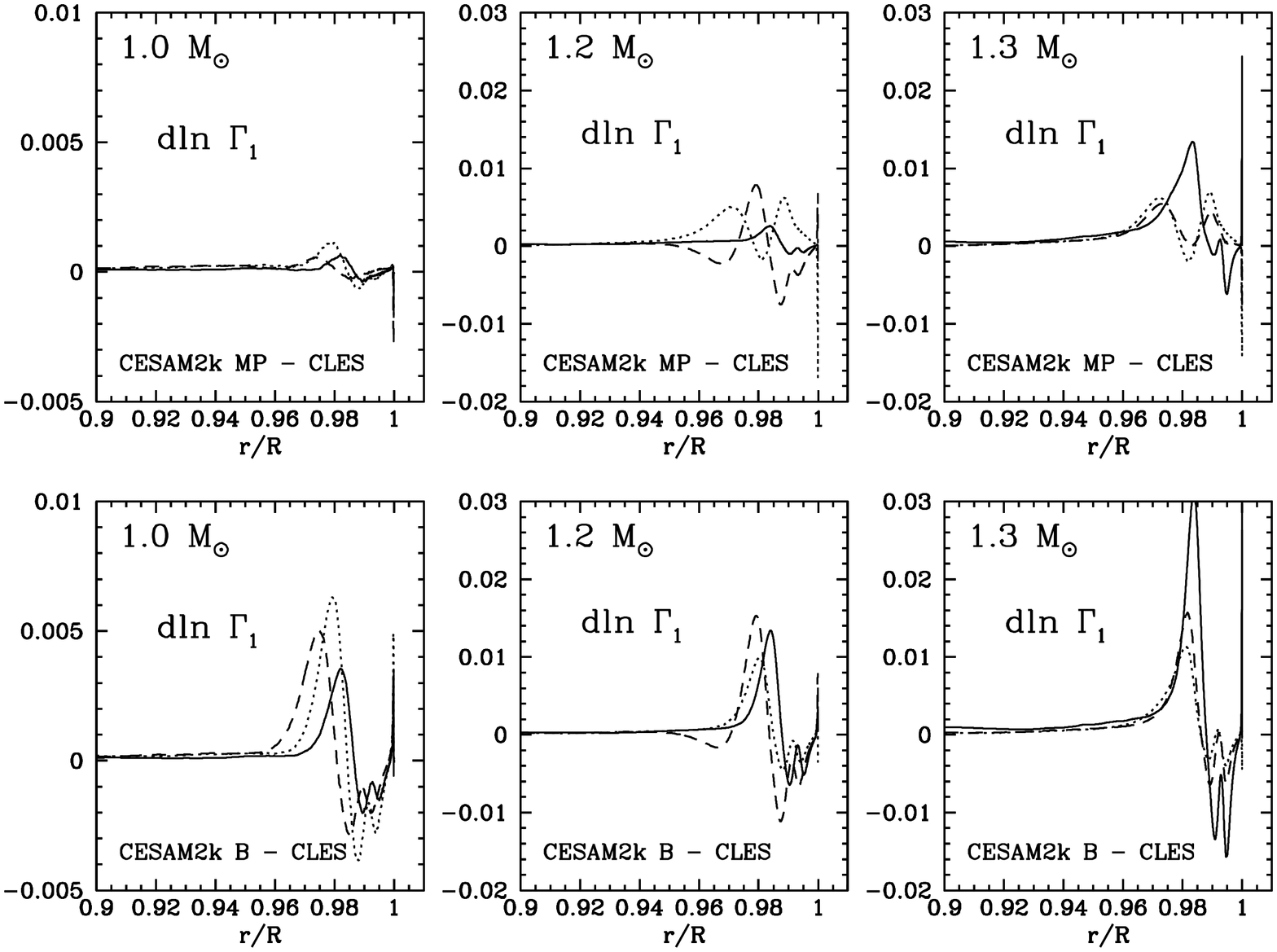}
\caption{Lagrangian differences of sound speed ($d\ln c$) (upper panels) and adiabatic
exponent ($d\ln\Gamma_1$) (lower panels) as a function of the normalised stellar radius for
models corresponding to task3.1 (left), task3.2 (centre), task3.3 (right). For each mass, 
three evolutionary stages are considered: A, with  $X_{\rm c}=0.35$ (solid lines), B, with
$X_{\rm c}=0.01$ (dotted lines), and C, with $M_c^{\rm He}=0.05 M_{\star}$ (dashed lines).}
\label{figdif}
\end{center}
\end{figure}

The internal structure at the given stages A, B and C can be studied by means of
 the sound speed, $c$, and of the adiabatic exponent, $\Gamma_1$, variations.
The Lagrangian differences, $d\ln c$ and $d\ln \Gamma_1$, between CESAM2k (both B and MP) and CLES
models (calculated at the same mass by using the
ADIPLS package tools\footnote{http://astro.phys.au.dk/~jdc/adipack.n})  
are plotted in Fig.~\ref{figdif} as a function of the normalised radius.
Note that the vertical scale in $d\ln c$ and $d\ln\Gamma_1$ plots are respectively 
five and three times smaller  for 1.0~\msun\  than for 1.2 and 1.3~\msun.
The $d\ln c$ values reflect: {\rm i}) the differences in  stellar radius (note that the largest values are
reached in  Task3.2 B CLES-CESAM2k~MP comparison, for which dlnR is of the order of 0.01);
{\em ii})  the different chemical composition gradients below the convective envelope
(features between $r/R=0.6$ and 0.8), as well as differences in the location of convection region
boundaries (at $r/R\sim 0.05$ for the convective core in Task3.2 and 3.3).

\begin{figure}[b]
\begin{center}
\vspace*{-1cm}
\resizebox{\hsize}{!}{\includegraphics{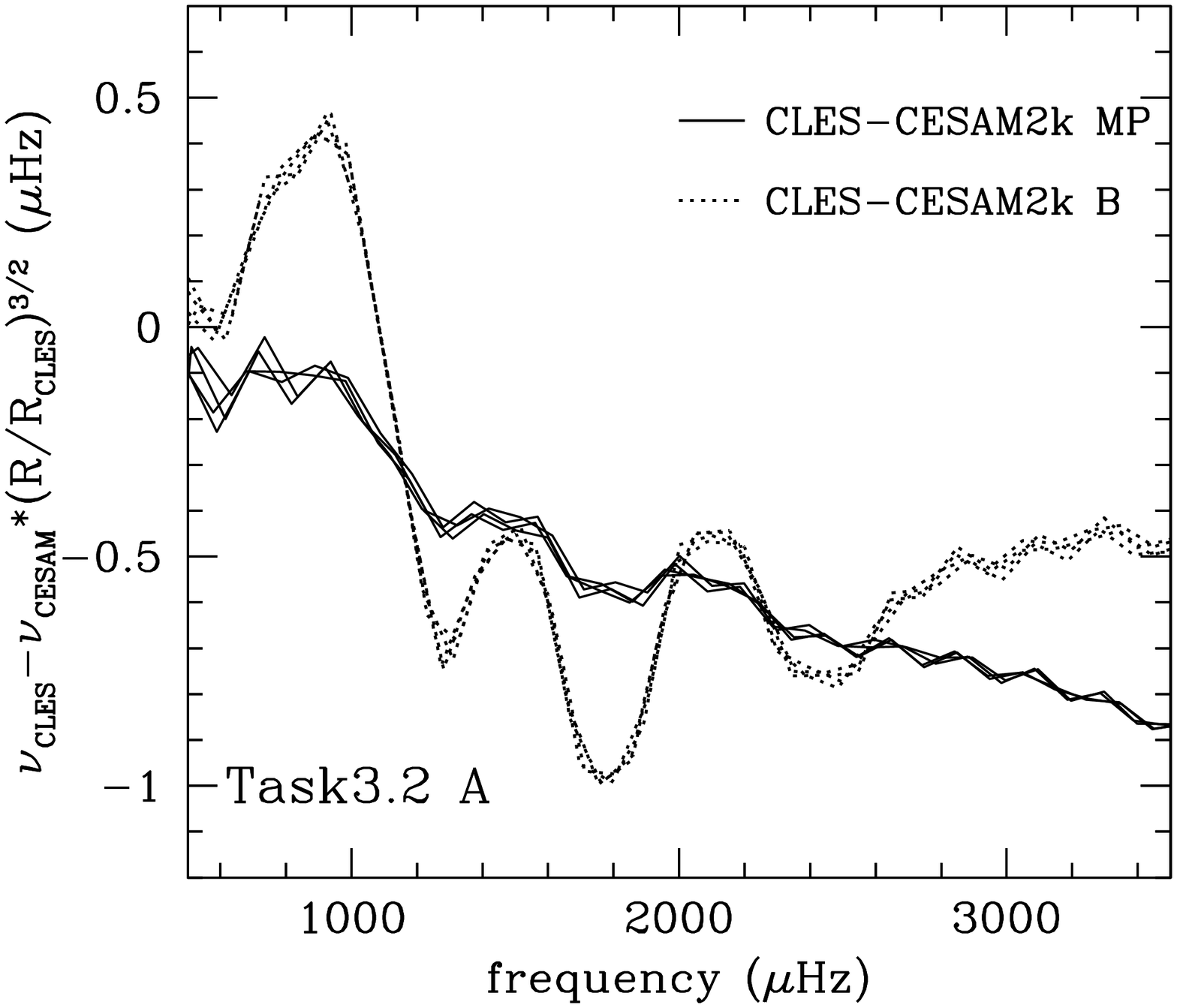}\includegraphics{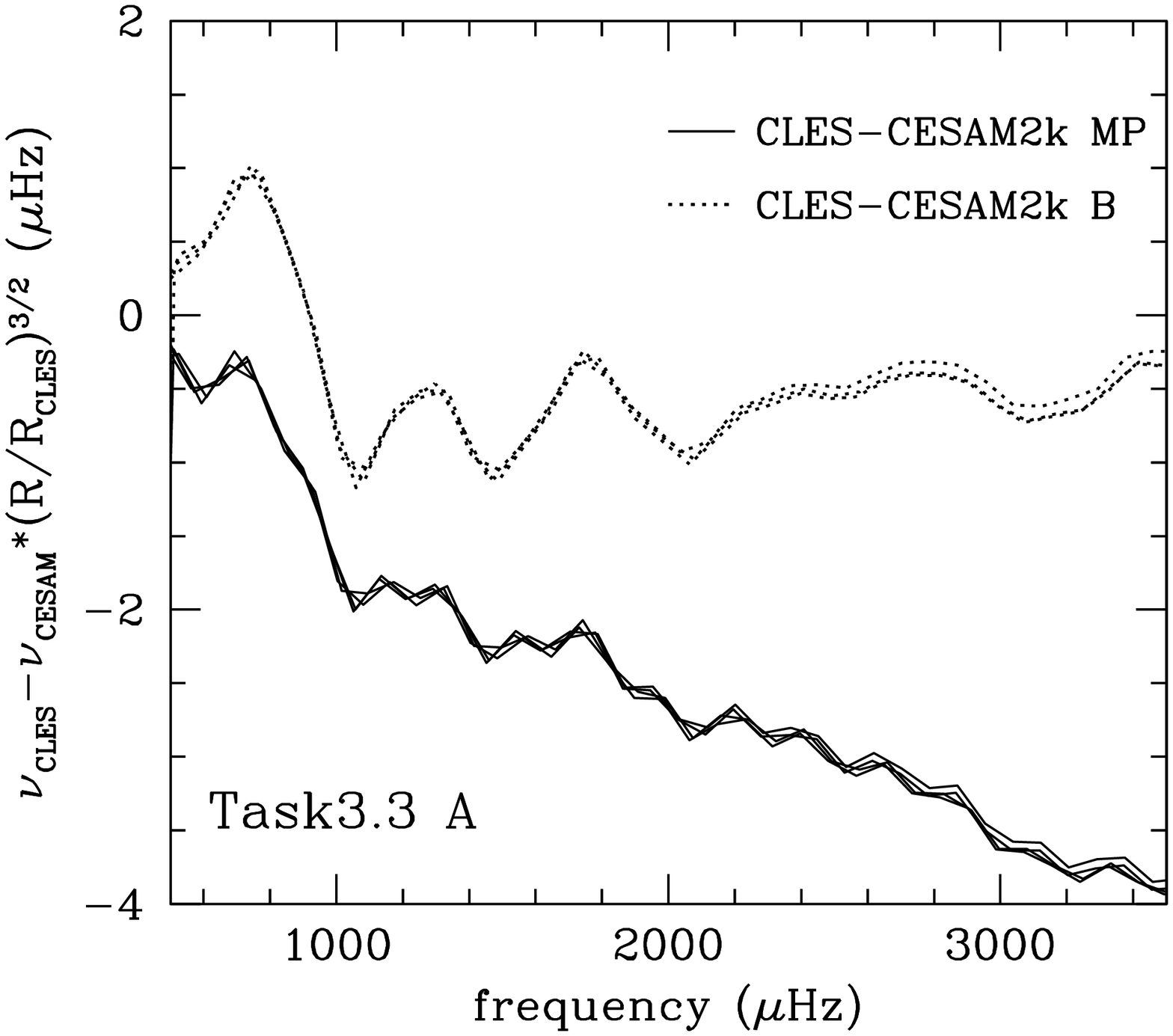}}
\caption{Plots of the frequency differences after removing the scaling due to stellar 
radius, between models computed
with CLES and CESAM2k~B (dotted lines) and with CLES and  CESAM2k~MP (solid lines).
 For each couple CLES-CESAM there are
four curves that correspond to different degrees $\ell=0,1,2,3$. 
Left panel: 1.2~\msun\ models with $X_{\rm c}=0.35$
Right panel: like left panel for 1.3~\msun\ models. 
}
\label{figfreq}
\end{center}
\end{figure}

The value of $\Gamma_1$ in the external regions is particularly sensitive to the He abundance.
Therefore, as one can see  in the bottom panels of Fig.~\ref{figdif},  the variations $d\ln\Gamma_1$ 
are  smaller for  CESAM2k~MP--CLES  comparisons than for CESAM2k~B--CLES ones, and 
these differences increase with the mass of the model; these results are in good agreement with 
what we would expect from $Y_{\rm S}$ curves in Fig.~\ref{figys}.

To clarify how all these differences affect the seismic  properties of the models, we 
compute by means of the adiabatic seismic code LOSC (Scuflaire {\em et al.}, 2007b) the frequencies
of oscillations of all the models at the evolutionary stage A (main sequence models).
In Fig.~\ref{figfreq} the frequency differences between CLES and CESAM models of 1.2 (left panel) and
1.3~\msun\ (right panel) are shown for p-modes with degrees $\ell=$0, 1, 2, 3.
The similar behaviour of curves with different degree indicates that the observed
frequency differences reflect mainly the near surface difference of the models.
In particular, the oscillatory component in CLES-CESAM2k~B  frequency differences is
the characteristic signature of the different He content in the convective envelope.
Note that the vertical  scale in both panels is not the same, and that the amplitude of the 
oscillatory component is related to the difference of surface He content.
Comparisons for 1.0~\msun\ models showed frequency differences of about 0.4~$\mu$Hz.

\begin{figure}
\begin{center}
\vspace*{-0.75cm}
\includegraphics[width=\textwidth]{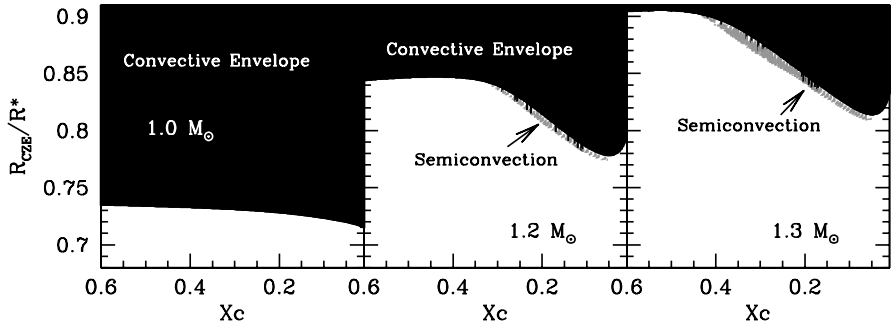}
\vspace*{-1cm}
\caption{Evolution of the radius of the convective envelope for 1.0 (left), 1.2 (middle) 
and 1.3~\msun\ (right). Black regions represent the convective envelope, and grey ones
the ``semiconvection'' region below the convective envelope.}
\label{semicon}
\end{center}
\end{figure}

\section{Boundaries of the convective regions}
  The evolution of the convective region boundaries   in models with metal diffusion
is  difficult to study.  
In fact, as it was already noted by Bahcall {\em et al.} (2001) in the case of
 1.0~\msun\ models, the accumulation of metals below the convective envelope 
 can trigger the onset of semiconvection. As the metal abundance increases below the 
convection region, the opacity locally increases 
and the affected layers end up by becoming 
convectively unstable\footnote{We recall that we use the classical Schwarzschild
criterion for convective instability}. 
\begin{figure}
\begin{center}
\includegraphics[scale=0.4]{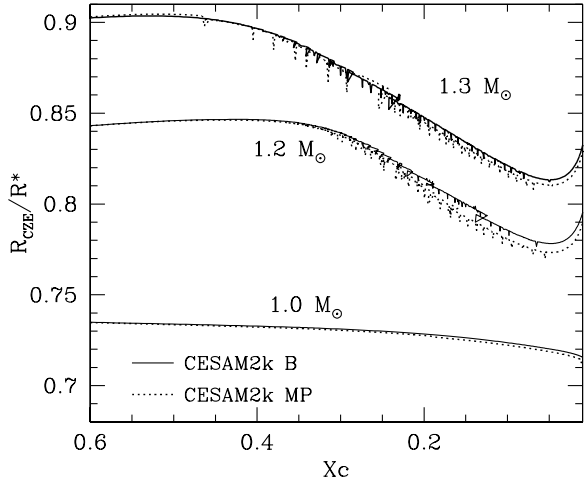}
\caption{Evolution of the radius of the convective envelope for 1.0, 1.2, 
and 1.3~\msun\ CESAM models. Solid lines correspond to CESAM2k with Burgers equations, and
dotted lines to CESAM2k with Michaud \& Proffitt (1993).}
\label{cecesam}
\end{center}
\end{figure}
The evolution of these unstable layers strongly depends on the numerical treatment of convection
borders used in the stellar evolution code.
CLES does not treat  semiconvection, and the algorithm computing the chemical
composition  in convective regions includes a kind of ``numerical diffusion''.
In CLES, the convectively unstable shells may grow
and eventually join the convective
envelope. As a consequence, the latter becomes suddenly deeper, destroys the Z gradient, recedes, 
and the process starts again. 
So, the crinkled profiles of $Y_{\rm S}$ for Task3.2 and 3.3 are a consequence of the 
sudden variations of the depth of the convective envelope. 
Since the timescale of diffusion decreases as the mass of the convective envelope decreases, 
semiconvection appears earlier in  1.3~\msun\ than in 1.2~\msun\ models. Furthermore, 
in contrast with Bahcall {\em et al.} (2001) results, semiconvection does not appear
in our evolved 1.0~\msun\ models, probably because of the effect of ``numerical diffusion'' that
reduces the efficiency of metal accumulation.  
All these effects can be seen in Fig.~\ref{semicon}. In Fig.~\ref{cecesam} we plot the evolution of
the convective envelope for CESAM models. The different treatment of convection borders in both
codes leads to different depth of the convective envelope. At $X_{\rm c}=0.05$,
CLES models have convective envelopes of about 0.1\% deeper than CESAM2k~B ones, and
of about 2.3\%, 0.6\% and 0.4\% shallower than CESAM2k~MP models for 1.0, 1.2 and 1.3~\msun\ 
respectively.  

\begin{figure}[t]
\begin{center}
\vspace*{-1cm}
\resizebox{\hsize}{!}{\includegraphics{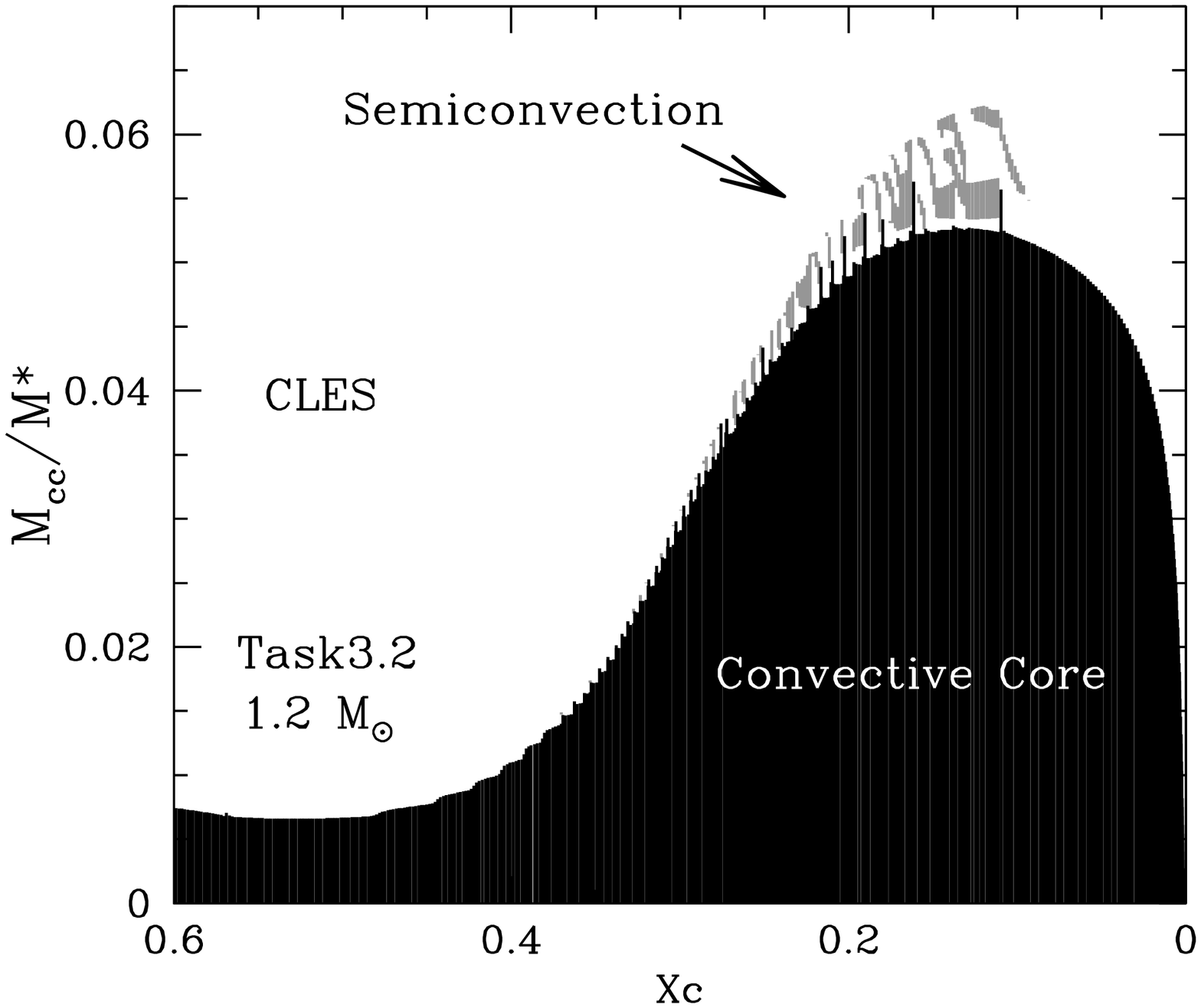}\includegraphics{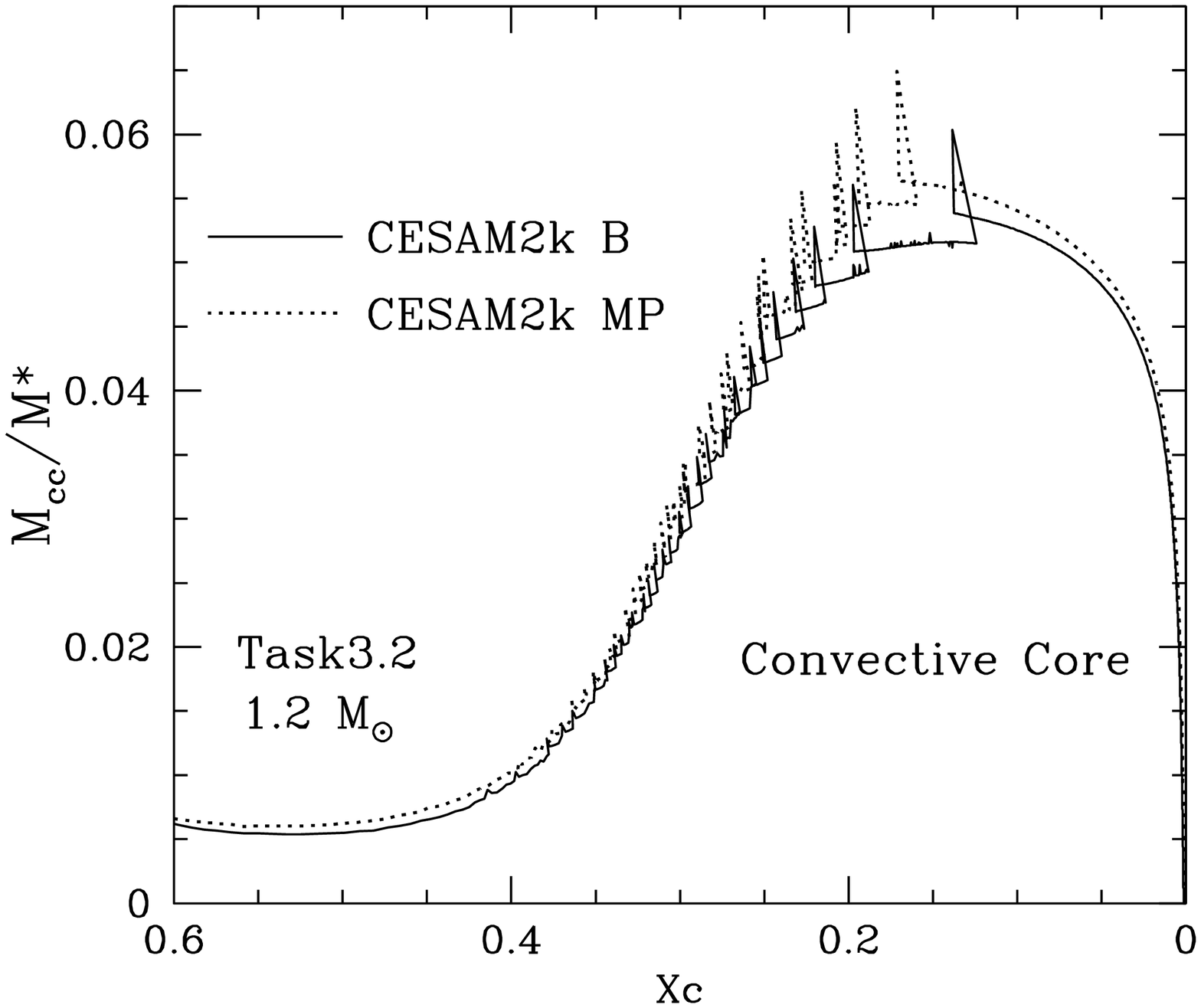}}
\caption{Convective core mass evolution  for 1.2~\msun\ evolution computed with CLES
(left panel) and with CESAM (right panel). 
The black region in left panel correspond to the convective core , and grey ones
are convectively unstable regions outside the convective core.}
\label{semicon_core}
\end{center}
\end{figure}

Semiconvection can also appear at the border of the convective core. As explained in Richard {\em et al.}
(2001), because of the He abundance gradient generated  at the border of the convective
core by nuclear burning, the diffusion term due to the composition gradient
counteracts the He settling term  and He ends up by going out of the core. Since the outward He 
flux interacts also with the metals, these may as well diffuse 
outward the core and prevent the metals settling. The enhancement of metals at the border of the 
convective core induces an increase in opacity and, finally, the onset of  semiconvection.
For the masses considered in Task3.2 and Task3.3, semiconvection appears very easily, as 
the mass of the convective core increases with time,
leading to a  quasi--discontinuity in the He abundance. 

As for the convective envelope, the numerical treatment of the border of the 
convective regions  is a key aspect of the convective core evolution. 
In Fig.~\ref{semicon_core} we plot the evolution of the convective regions in the central part of 
1.2~\msun\ models computed with CLES (left panel) and with CESAM (right panel).
While CLES treatment of convective borders  keeps convectively unstable shells 
separated from the convective core (grey region), it seems that CESAM tends to connect these
shells to the central convective region. In fact, the envelope of the curve M$_{cc}$ {\rm vs.} $X_{\rm c}$
for CESAM2k~MB model approximately coincides with the ``semiconvection'' region in CLES plot.
As a consequence, a larger central mixed region in CESAM2k~MP than in CLES leads to a longer
main sequence phase, as seen in Fig.~\ref{fighr}.
In fact, the value of $M_{\rm cc}$, just  before it begins to decrease, is  
6\% and 12\%, respectively for 1.2 and 1.3~\msun, larger for CESAM2k~MB models
than for CLES ones. On the other hand, the corresponding values for CESAM2k~B are
2\% and 10\% larger than CLES ones.

\section{Diffusion coefficient differences}

\begin{figure}[b]
\begin{center}
\vspace*{-1cm}
\resizebox{\hsize}{!}{\includegraphics{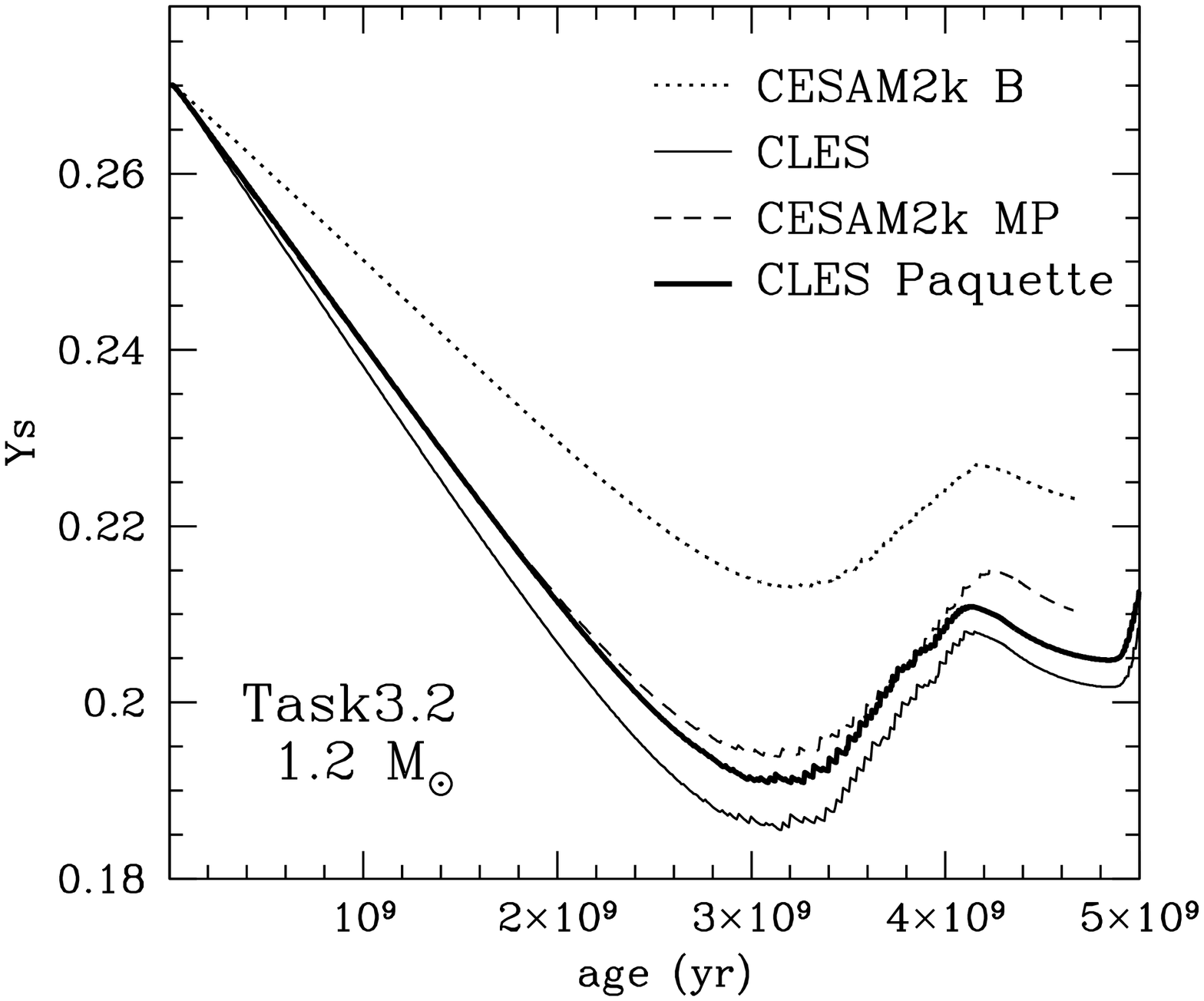}\includegraphics{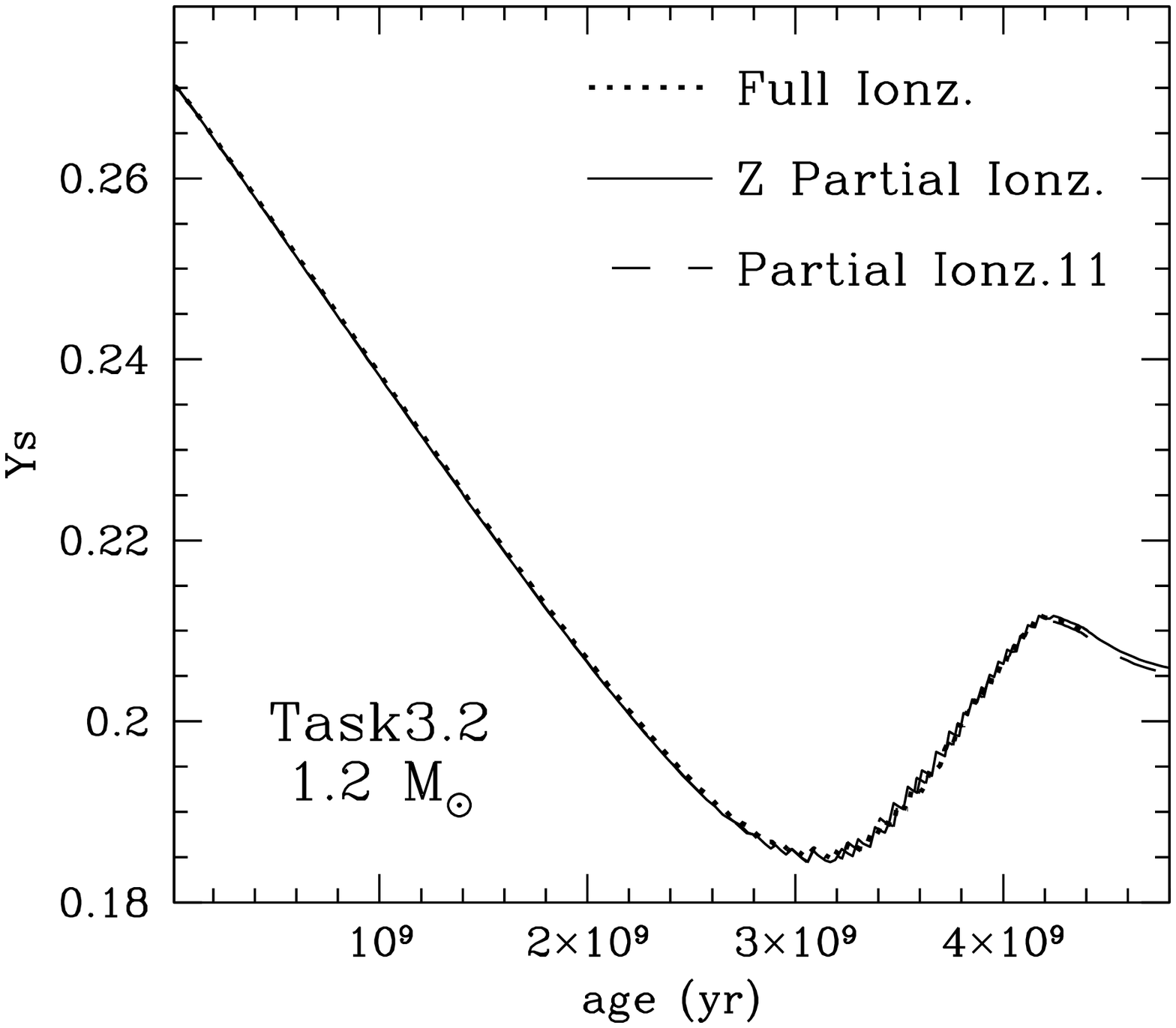}}
\caption{Curves of the helium mass fraction in the
convective envelope as a function of age for a 1.2\msun\ evolution.
Left panel: thick line corresponds to models computed by CLES 
with Paquette coefficients, and the other three curves corresponds
to the results already shown in Fig.~\ref{figys}.
Right panel: evolution of the helium mass fraction in the convective envelope for
Task3.2 models computed by an updated version of CLES assuming: full ionization (dotted line),
partial ionization of the ``average'' element Z (solid line) and partial ionization of eleven 
elements that diffuse separately (dashed line).}
\label{yspaquette}
\end{center}
\end{figure}

The discrepancies we found between CESAM2k~MP and CLES diffusion efficiency are in good agreement
with the comparisons already published by TBL94. 
The large differences between CESAM2k~B and CLES are instead rather unexpected. 
Both codes, in fact,  derive the diffusion velocities by solving the  Burgers' equations,
however, the values of friction coefficients appearing in those equations are different in
CESAM2k~B and CLES approaches. The resistance coefficients $K_{\rm ij}$, which represent
the effects of collisions between the particles i and j, are
$K_{\rm ij}=C_{\rm ij}\,F_{\rm ij}^{(11)}$ in CESAM2k~B, and
$K_{\rm ij}=C_{\rm ij}\,2\,\ln \Lambda_{\rm ij}$ in CLES (TBL94).
The term $C_{\rm ij}$ is the same 
in both formulations and depends on the mass, charge and concentration of the particles i and j.
The values of the quantity $F_{\rm ij}^{(11)}$ are derived from the numerical fits of the
collision integrals (Paquette {\em et al.},1986), and the 
term $\ln \Lambda_{\rm ij}$ is the Coulomb logarithm from Iben \& MacDonald (1985).
Furthermore, while TBL94
adopt for the heat flux terms $z_{\rm ij}$, $z'_{\rm ij}$ and $z''_{\rm ij}$ their
 low density asymptotic values, CESAM2k~B computes them by using the
collision integrals from  Paquette {\em et al.} (1986).

As shown in Thoul \& Montalb\'an (2007), 
the assumptions done in TBL94 can lead, for the Task3.2~A model,
to diffusion velocities between 6 and 20\% larger than those that would be
obtained by using the Paquette's coefficients. 
To further clarify this point, we replaced in Burgers equations the coefficients 
used in CLES with those used in CESAM2k~B and 
we re-computed  the models for Task3.2. The new evolution of He surface abundance
is plotted in Fig.~\ref{yspaquette} (left panel, thick line)
together with the curves obtained directly by CESAM2k~B, standard CLES, and CESAM2k~MP.
We see that the approximation adopted in TBL94 implies a helium surface
abundances slightly smaller than those that would be obtained by using the numerical
fits by Paquette. The new CLES values are close to CESAM2k~MP ones, but 
still quite far from CESAM2k~B results.

 Another important difference between CESAM and CLES diffusion routines is
that, while CESAM follows separately each element inside Z  and
determine the ionization degree of all the species, the standard version of CLES adopts 
 full ionization, and  follows only four species: H, He , electrons and an ``average'' 
element Z with atomic mass 8, charge 17.84.
To test the consequences of these approximations we  computed the evolution of 1.2~\msun\
with an updated version of CLES that computes the ionization degree, and allows to follow
separately up to 22 elements. In Fig.~\ref{yspaquette} (right panel) 
we plot the evolution of the He surface abundance for calculations considering full ionization, 
and partial ionization for the eleven most relevant elements in Z. 
 We can conclude that, at least for masses lower than or
equal to 1.2~\msun\, the effect of partial ionization on the He diffusion velocity is negligible.

Finally, we  checked the effect of the time step by computing CLES evolution tracks
with smaller and larger steps, but no significant effect was detected in the
diffusion efficiency.

\section{Conclusions}        
We  compared models corresponding to Task3.1, Task3.2 and Task3.3 which were computed with
three different implementations of microscopic diffusion.
The largest discrepancy ($\sim$40\%) appears between codes that model diffusion velocities by solving
the Burgers' equations (CESAM2k~B and CLES). A detailed analysis  showed  that the approximations
used in Thoul et al. 1994 for the friction coefficients are not at the origin of this
discrepancy. 
Computations with partial ionization has also shown that for masses smaller or equal to 1.2\msun,
the full ionization assumption has no detectable effects.
Therefore, we conclude that the  difference between CLES and CESAM2k~B results
origins  from the routine solving the Burgers' equation system.

Moreover, we showed that the effect of the different treatments of the convection borders 
can lead, when diffusion is included,  to significant discrepancies (up to 12\%) 
for the mass and radius of convective regions.

\section*{Acknowledgments}
The authors thank HELAS for financial support. JM and ST are supported by Prodex 8 COROT (C90197)

\end{document}